\documentclass[english,twocolumn,aps,prl, superscriptaddress,showpacs, amsmath]{revtex4-1}
\usepackage{graphicx}

\usepackage{color}

\usepackage{natbib}
\usepackage[T1]{fontenc}
\usepackage[latin9]{inputenc}
\usepackage{color}
\usepackage{units}
\usepackage{amssymb}
\usepackage{graphicx}
\usepackage{esint}
\usepackage{bm}
\usepackage{natbib}
\usepackage{ulem}
\usepackage[colorlinks=true, citecolor=blue]{hyperref}
\setcitestyle{journalcolor= blue}

\begin{document}

	\title{Superconducting Dirac point in proximetized graphene}

	\author{Gopi Nath Daptary}
	\author{E. Walach}
	\author{E. Shimshoni}
	\author{A. Frydman}
	
	\affiliation{Department of Physics, Jack and Pearl Resnick Institute and the Institute of Nanotechnology and Advanced Materials,
		Bar-Ilan University, Ramat-Gan 52900, Israel}

\date{\today}

\begin{abstract}

  \textbf{Two-dimensional (2D) materials, composed of single atomic layers, have attracted vast research interest since the breakthrough discovery of graphene. One major benefit of such systems is the simple ability to tune the chemical potential by back-gating, in-principle enabling to vary the Fermi level through the charge neutrality point, thus tuning between electron and hole doping. For 2D Superconductors, this means that one may potentially achieve the strongly-coupled superconducting regime described by Bose Einstein Condensation physics of small bosonic tightly bound electron pairs. Furthermore, it should be possible to access both electron and hole based superconductivity in a single system. However, in most 2D materials, an insulating gap opens up around the charge neutrality point, thus preventing approach to this regime. Graphene is unique  in this sense since it is a true semi-metal in which the un-gapped Dirac point is protected by the symmetries. In this work we show that single layer graphene, in which superconducting pairing is induced by proximity to regions of a low density superconductor, can be tuned from hole to electron superconductivity through the strong coupling regime. We study, both experimentally and theoretically, the vicinity of this "Superconducting Dirac point" and find  an unusual  situation  where reflections at interfaces between normal and superconducting regions within the graphene, suppress the conductance and, at the same time,  Andreev reflections maintain a large phase breaking length. In addition, the Fermi level can be adjusted so that the momentum in the normal and superconducting regimes perfectly match giving rise to ideal Andreev reflection processes. }

\end{abstract}


\maketitle

A key ingredient in the Bardeen-Cooper-Schrieffer (BCS) theory, which has been very successful in explaining weakly coupled superconductivity, is that $E_F >> \Delta$, where $E_F$ is the Fermi energy and $\Delta$ is the pairing energy gap; thus pairing is restricted to a narrow shell of width $\Delta/\hbar v_F$ near the Fermi sphere. In the other limit, strong coupling superconductors, the typical Cooper-pair size or coherence length, $\xi$, is of the order of the interpair distance $1/k_F$, or equivalently, the superconducting gap, $\Delta$, is of the order of the Fermi energy, $E_F$. This represents Bose-Einstein condensation (BEC) superconductivity of tightly bound Bosonic electron pairs.

\begin{figure*}
	\includegraphics[width=19cm,height=!]{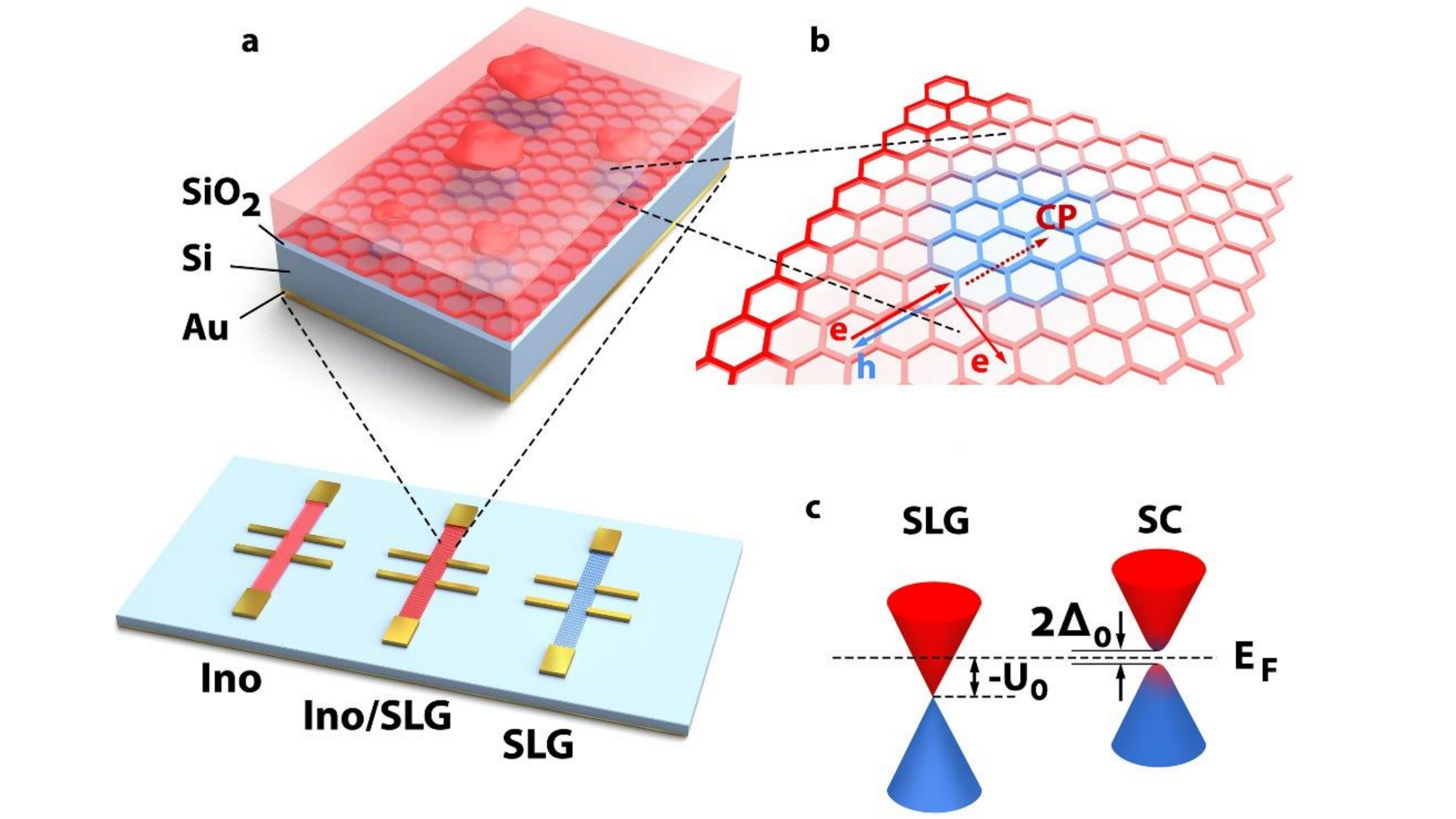}
		\caption{\label{setup}
	\textbf{Sketches of the Graphene/InO heterostructures. a,} Schematic diagram of the three devices (bare InO, bilayer of SLG/InO, and bare SLG) grown on a Si/SiO$_2$ substrate (bottom panel). A gold evaporated electrode on the bottom is used to apply a backgate voltage on the SLG enabling the tuning of $E_F$. A zoom of the SLG/InO (top panel) illustrates superconducting grains within the InO insulating layer. These induce proximity based regions of finite pairing (light blue) in the graphene layer.  \textbf{b,}  In the SLG an electron impinging on the NS-interface is reflected as a quasi-particle which has electron (e, red arrow) and hole (h, blue arrow) components  while transferring a cooper pair (CP) into the Superconductor. \textbf{c,} Band diagram of normal parts of the SLG (left) and the proximity induced superconducting puddles (right)
charecterized by a Fermi energy difference $U_0$ and an induced pairing gap $\Delta_0$, when tuned to the special point $E_F^\prime=E_F+U_0=0$
 (see text).}
\end{figure*}

The BCS-BEC crossover has been widely studied in cold atom systems in which the pairing gap is modified by tuning the attractive interaction \cite{stewart2008using,gaebler2010observation}. In superconductors, however,  $\xi$ is usually much larger than $1/k_F$ and BEC physics is not attainable mainly due to the fact that it is very difficult to control $\Delta$. In most strongly correlated superconductors such as Cuprates, some kind of insulator (typically a Mott insulator) is developed at zero doping and superconductivity appears only beyond a threshold value of doping. Exceptionally, a number of iron based superconductors are found to be located in the BCS-BEC crossover regime \cite{kasahara2016giant} especially in systems in which $E_F$ can be modified by chemical doping \cite{rinott2017tuning}.

Two dimensional superconductors, in which $E_F$ can be tuned by gating, are promising systems for obtaining the BCS-BEC crossover regime.
The rapid progress in fabrication of Van-der-Waals 2D materials \cite{novoselov20162d,ajayan2016van,duong2017van} suggests a variety of novel platforms. However, in most of these materials a band gap opens at the charge neutrality point, and superconductivity is manifested only at substantial carrier densities.
In this respect, Graphene is unique: it exemplifies a true semi-metal where un-gapped band-touchings (Dirac points in a single layer graphene (SLG)) at the two valleys are protected by the lattice symmetries, and by the tiny value of spin-orbit coupling \cite{neto2009electronic}. The problem is that in its natural form, graphene does not posses superconductivity either.

In this work we study a SLG in which a finite pairing gap is induced due to proximity to a low-density, strongly disordered superconductor. In such a system, not only is the BEC regime accessible, but it is also possible to tune between an electronic Cooper pair  to a hole Cooper pair based superconductor. Close to charge neutrality, a peculiar type of BEC-like state develops with a "conventional", s-wave pairing. In addition, this system can be tuned to an optimal doping where perfect Andreev reflections occur at normal-superconductor interfaces.

\begin{figure*}
	\includegraphics[width=18cm,height=!]{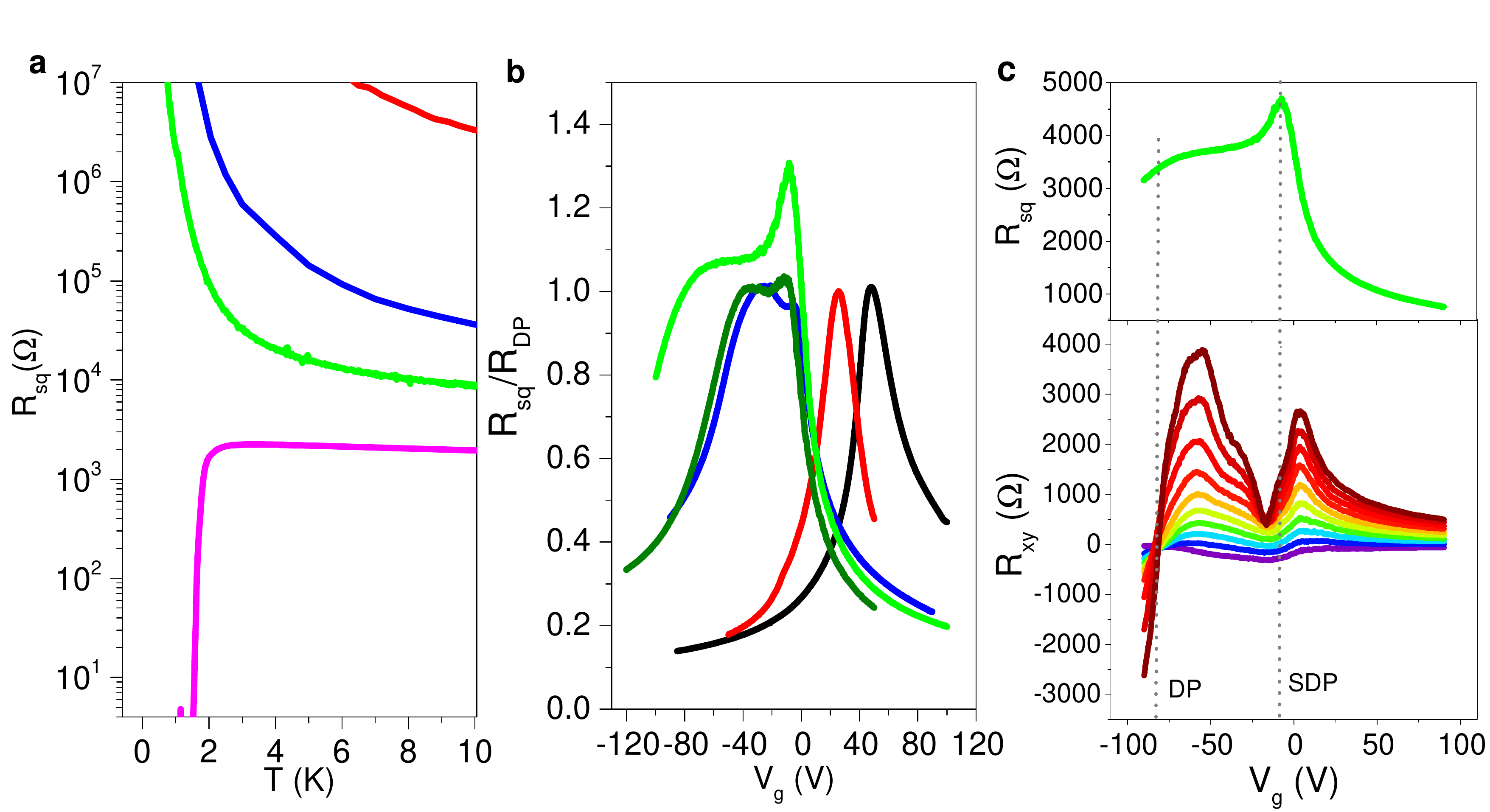}
	\caption{\label{Rxx-Rxy}
		\textbf{Sheet resistance and Hall resistance as a function of temperatures and gate voltages. a,} Sheet resistance $R_{sq}$ of a series of bare InO films as a function of temperature grown at different O$_2$ pressure, S1, S2, S4 and S5 from top to bottom respectively. \textbf{b,} Sheet resistance $R_{sq}$, normalized by the resistance at the Dirac point,  as a function of $V_g$ of a SLG (black solid line) and SLG/InO bilayer S1-S4 from left to right respectively (S5 is superconducting, and hence it shunts the SLG). Measurements were performed at $T = 1.7$ K and $B = 0$ Tesla.  \textbf{c,} Lower panel: Hall resistance, $R_{xy}$, of sample S4 as a function of $V_g$  at $T = 1.7$ K and different magnetic field $B = 0-9$ Tesla (in steps of 1 Tesla ). Note that the charge neutrality point is at $V_d = -80$ V. Upper panel: $R_{sq}$ of sample S4 as a function of $V_g$ at $T = 1.7$ K and $B = 0$ Tesla. }
\end{figure*}

The disordered superconductors in our experiments were 30 nm thick films of amorphous indium oxide (InO) that were e-beam evaporated on a patterned CVD grown SLG . The O$_2$ partial pressure during evaporation (in the range 2 - 8$\times$10$^{-5}$ Torr) determined the carrier density, $n$, of the disordered superconductor in the range $10^{19}-10^{20}$ cm$^{-3}$ \cite{ovadyahu1986some}, a few orders of magnitude smaller than typical $n$ in metals. This work includes five SLG/InO stacks in which this partial pressure was $8 \times 10^{-5}$ (S1), $6.2 \times 10^{-5}$ (S2), $4 \times 10^{-5}$ (annealed) (S3), $4 \times 10^{-5}$ (S4),  and $2 \times 10^{-5}$ (S5).  In all samples the InO was insulating except for S5 in which the InO was superconducting. For reference, we simultaneously prepared samples of bare SLG and InO (Fig \ref{setup}a).  As the $n$ is increased and $R_{sq}$ decreases, the InO films undergo a superconductor - insulator transition (SIT) as seen in Fig. \ref{Rxx-Rxy}a.

InO films, despite being morphologically uniform, have been shown to include emergent granularity in the form of superconducting puddles embedded in an insulating matrix \cite{kowal1994disorder}. Hence, local superconductivity is present even in the insulating phase of the SIT. This phase, in which local electronic pairing is present, has been dubbed a "Bosonic insulator".  Indeed, experiments have revealed evidence for superconducting vortices and a finite energy gap in the insulator phase of InO \cite{sacepe2011localization,poran2011disorder,sherman2012measurement,kopnov2012little,roy2018quantum}. The small superconducting puddles have a higher electron density than the insulating background. Coupling such a "superconducting insulator" to a SLG film gives rise to a unique situation. The underlying graphene develops regions with a non-vanishing superconducting gap due to proximity effect just below the superconducting puddles (Fig. \ref{setup}a). Depending on electrostatic details electron density in the proximatized puddles can be locally depleted or inflated relative to the normal graphene background. As a consequence, the puddles can be described by two quantities, $\Delta_0$ and $U_0$, describing the induced superconducting gap and the potential difference between the puddles and the insulating background respectively. It turns out that in our case, $U_0$ is negative (opposite to the case considered theoretically in the past \cite{beenakker2006specular}). as a result, in the S region the effective Fermi level $E'_F= E_F + U_0$  is lower than $E_F$ in the N region (see Fig. \ref{setup}c). Due to the low carrier density in the InO film, $U_0$ can be moderately small thus providing the experimental opportunity to tune the chemical potential through the charge neutrality point of both the normal regions and the proximity induced superconducting islands. In the remainder of this paper, the former is named the Dirac point (DP) and the latter is dubbed the superconducting Dirac point (SDP).

The presence of two such points can be traced by measuring the bilayer resistance as a function of gate voltage, $V_g$, which controls $E_F$. Figure \ref{Rxx-Rxy}b shows $R(V_g$) for a SLG and a series of SLG/InO bilayers (samples S1-S4). For the bare SLG, the DP occurs at a gate voltage $V_g \approx 50$ V indicating that the graphene is hole doped due to adsorption of atmospheric dopants such as H$_2$O and O$_2$ \cite{shin2010surface}. As the InO film is tuned towards the SIT by increasing the carrier density, an increasing electron doping is induced in the graphene layer so its DP shifts to negative gate voltages. At the same time, an additional resistance peak emerges to the right of the DP and grows with closeness of the InO film to the SIT. This second resistance peak is related to superconductivity and is identified as the SDP. Hall effect measurement performed on the bilayer samples (see Fig. \ref{Rxx-Rxy}c) reveal that the DP is indeed the global point of charge neutrality. However, the SDP also has a very distinct signature on the the $R_{xy}(V_g)$ curves and one can envision that it superimposes an additional charge neutrality feature on the overall Dirac behavior background.

\begin{figure*}
	\vspace{0.5cm}
	\includegraphics[width=17cm,height=!]{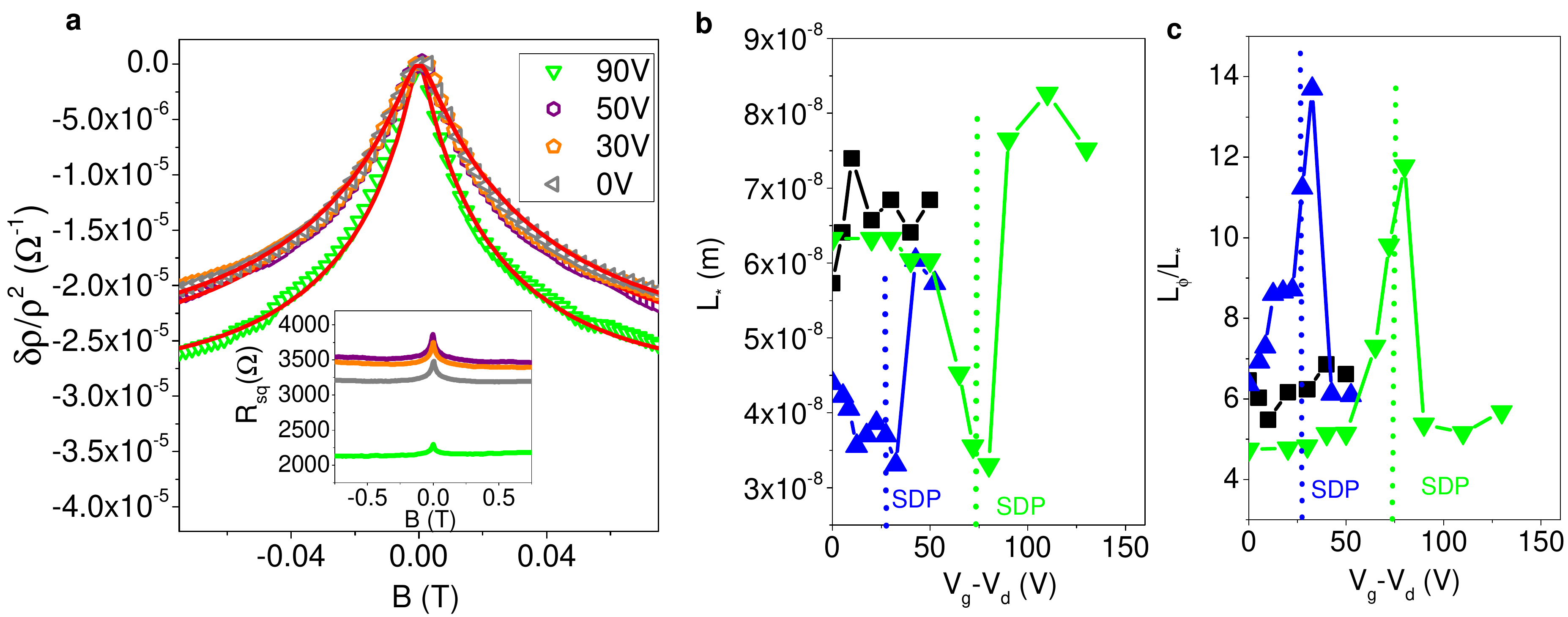}
	\caption{\label{WL}
		\textbf{Weak localization a,} Relative sheet resistance, $\delta \rho/\rho^2$,  versus magnetic field as a function of $V_g-V_d$ of sample S4 including fits to Eq. 1 in SI. Raw data of these curves is shown in the inset. \textbf{b,}   $L_*$ and \textbf{c,} $L_\phi/L_*$  extracted from these fits as a function of $V_g-V_d$ for SLG (black), S2 (blue) and S4 (green). Measurements were performed at $T$= 1.7~K. }
\end{figure*}

The SDP has also a distinct signature on the weak localization (WL) contribution to the conductivity. Figure \ref{WL}a shows the resistance as a function of field for sample S4 at different values of $V_g$ and $T = 1.7$ K, exhibiting WL at the low magnetic field regime. From  these curves we extract the values of the phase breaking length, $L_\phi$, the inter-valley elastic scattering length,  $L_i$, and the intra-valley elastic scattering length,  $L_*$ \cite{mccann2006weak,tikhonenko2008weak} (see Supplementary). $L_*$ as a function of $V_g$ shown in Fig. \ref{WL}b  reveals a clear minimum at the SDP, implying that {\em intravalley} scattering is greatly enhanced at this point. In contrast,  $L_i$ is relatively constant irrespective of the InO details or gate voltage as shown in Supplementary. $L_\phi / L_*$, which determines the magnitude of the WL \cite{tikhonenko2009transition} peaks at the SDP as shown in Fig. \ref{WL}c.

\begin{figure*}
\includegraphics[width=12cm,height=!]{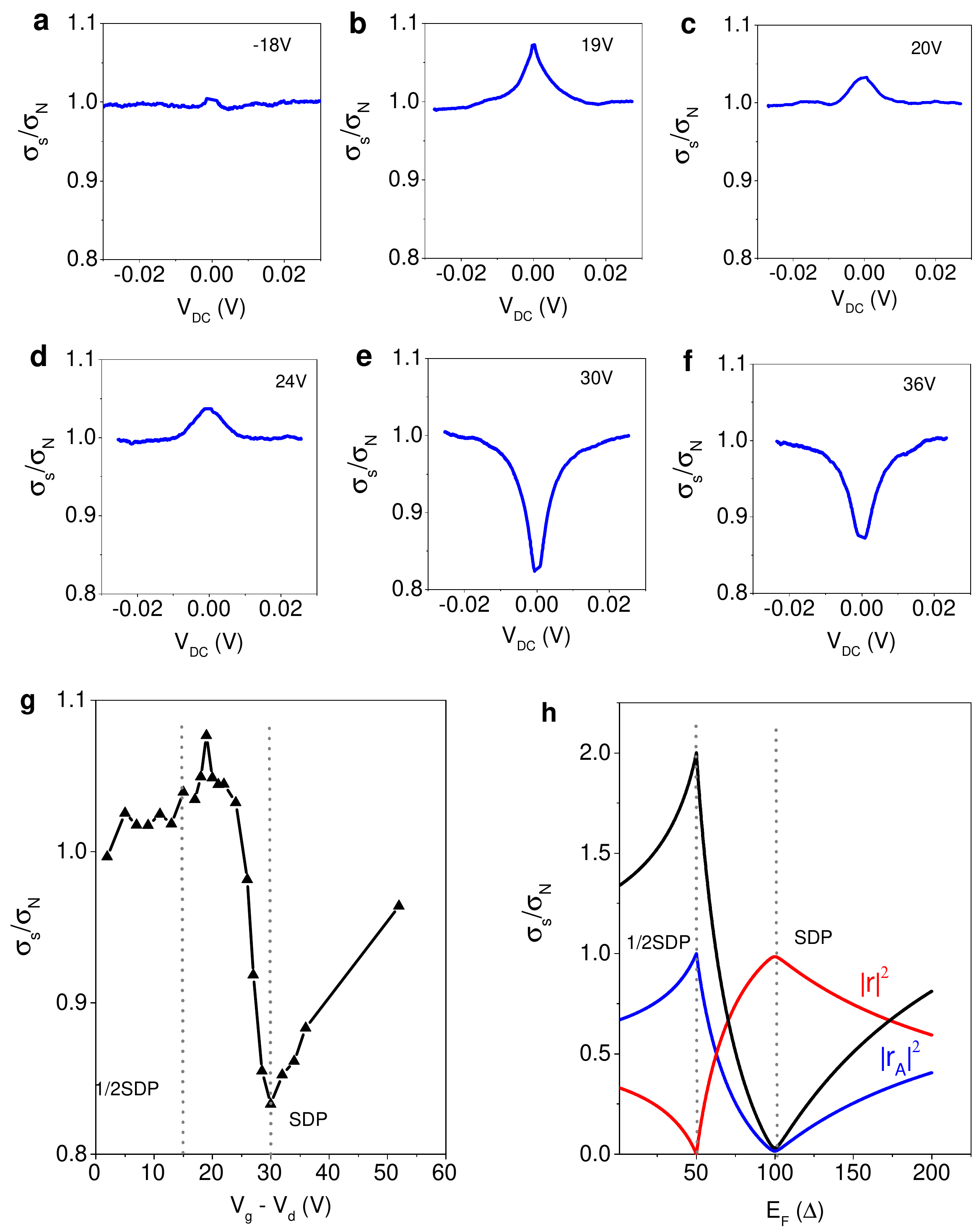}
\caption{\label{dIdV}
\textbf{Normalized conductance versus of bias voltage. a-f,} Relative conductivity $\sigma_s/\sigma_N$ ( $\sigma_s$ is the differential conductivity of bilayer of SLG/InO and $\sigma_N$ is the differential conductivity of bare SLG) as a function of bias voltage $V_{DC}$  of sample S3. \textbf{g,} The extracted $\sigma_s/\sigma_N$ at  $V_{DC}=0$ as a function of $V_g-V_d$ at $T = 0.33$ K. \textbf{h,} $\sigma_s/\sigma_N$ (black line) resulting from the theoretical model as a function of $E_F$ in units of $\Delta_0$; the normal and Andreev reflection rates ($|r|^2$, $|r_A|^2$) are indicated in red and blue, respectively.}
\end{figure*}

The behavior of $L_{\phi,i,*}$ are understood in the context of the different scattering processes on the interface between a normal region, N, and a proximity induced puddle, S, within the graphene film (Fig. \ref{setup}b). We consider an electron impinging from the N side at an incidence angle
$\alpha$ from the normal to the NS-interface, colliding with the interface, and then reflecting  as a quasi-particle which has electron and hole components.  The reflection process splits into a regular reflection (as an electron obeying the mirror reflection law) at amplitude $r$, and an Andreev reflection (AR) whereby a hole is reflected at amplitude $r_A$. One distinguishes two types of AR processes: retro-reflection, where the hole is scattered exactly backwards (while creating a Cooper pair in the S region), or a specular reflection where the hole is mirror-reflected at the interface. Notably, the latter type exists only very close to the Dirac point where $E_F < \Delta_0$. It is also important to note that while the reflected hole is in the opposite valley, the reflected electron is always in the same valley (intravalley scattering).

We evaluate $r$ and $r_A$ and derive transport properties using the Dirac-Bogoliubov-de-Gennes equation and the Blonder-Tinkham-Klapwijk formula \cite{blonder1982transition} for the conductivity, $\sigma_s=dI/dV$, at the interface (see Supplementary). The main results are presented in Fig. \ref{dIdV}h. We find that as $E_F$ is tuned by $V_g$, the behavior of $\sigma_s$ evolves between two extremes, and two special values of $E_F$ are identified:

\begin{enumerate}
	
	\item {The first special point is $E_F = -U_0$. This corresponds to $E'_F = 0$, where the superconducting regions are at the charge neutrality point. This SDP is quite unique: in its vicinity (i.e., for $E'_F< \Delta_0$), the electronic state in the S regions behaves effectively as a strongly-coupled superconductor, thus realizing the BEC limit. Since quasi-particle excitations of this state are strongly localized near the interface, one gets a perfect reflection ($\lvert r \rvert^2 = 1)$ at almost any angle $\alpha$. This yields a sharp peak in the resistance (Fig. \ref{Rxx-Rxy}b) and a minimum of the intra-valley scattering length,  $L_*$, as observed in Fig. \ref{WL}b.  However,  for $\alpha \rightarrow 0$,  i.e. for electron-states propagating almost perpendicularly to the interface, $r_A \rightarrow 1$. Namely, despite the small value of
$E'_F$, at normal incidence the AR prevails just the same as in more standard graphene NS-interfaces \cite{beenakker2006specular}. This behavior is a consequence of conservation of chirality  since for $\alpha=0$ the incident electron and reflected hole come from the same sub-lattice (similar to the situation of the Klein paradox \cite{katsnelson2006chiral}). For this reason, the NS-interface acts on one hand as a strong back-scatterer, but on the other hand contributes to a slight enhancement of $L_\phi$ compared to the bare SLG (see Supplementary). The net effect is the sharp peak in $L_\phi/L_*$ indicated
in Fig. \ref{WL}c.}
	
	\item {Another interesting point is $E_F = -\frac{U_0}{2}$. At this point, $\lvert E'_F \rvert = E_F$
		and there is perfect Fermi-momentum matching which allows ideal electron-hole transmission. As a result, $r$ and $r_A$ exhibit the opposite trend compared to the previous case: $\lvert r_A \rvert = 1$ at almost any $\alpha$, except for almost parallel incident waves  $\alpha \approx \pi /2$. The relative conductance is therefore close to twice the normal one.}
	
\end{enumerate}
It should be noted that for arbitrary value of $E_F$, the NS-interface contributes nothing to regular inter-valley scattering processes since the reflected electron is always in the same valley. This is consistent with the observed insensitivity of $L_i$ to variations in the InO sample details or the gate voltage (see Supplementary).

For characterizing the AR and studying these two unique points we performed differential conductance versus bias voltage measurements at different $V_g$ for sample S3. The results are summarized in Fig \ref{dIdV}. In order to avoid WL effects the measurements were performed at $B=6$ Tesla and the resulting traces of quantum oscillations were subtracted from the main feature. It is seen that for low gate voltages the zero bias differential conductance, ZBC, shows a maximum, typical of large Andereev reflection. This feature peaks at $V_g \approx 0.5V_{SDP}$. Clearly, $\sigma_S / \sigma_N$ is much smaller than the factor 2 which is expected for $r_A=1$. In addition, the voltage scale of the $\sigma(V_{DC})$ feature seen in Fig. \ref{dIdV}a-f  is an order of magnitude larger than $\Delta$ ($\approx 0.7mV$ \cite{sherman2012measurement}). These are due the fact that our sample contains many SN interfaces in series. As $V_g$ is increased, the ZBC decreases rapidly and shows a dip, indicating the suppression of $r_A$ and a rapid increase of $r$ which peaks at the SDP (Fig. \ref{dIdV}g). For voltages higher than $V_{SDP}$ the ZBC gradually increases towards the normal value. These results are in qualitative agreement with the theoretical traces presented in Fig. \ref{dIdV}h.

In summary, we show that graphene proximatized by a low density superconductor effectively exhibits strong coupling superconductivity, which can be tuned from hole to electron superconductivity. For our SLG/InO bilayers, in the gating range between the DP and SDP, the S regions are hole-doped while the N regions are electron doped. This gives rise to a wide range of parameters where one may realize an interface supporting Majorana modes and chiral Majorana fermions \cite{zhao2020interference}, which provides a new platform for topological quantum computing \cite{lian2018topological}. Furthermore,  chiral Andreev edge states may form in the integer quantum Hall regime. Indeed a study of our bilayers at high magnetic fields, potentially exhibiting such states, is currently being performed.

As noted, the samples studied in this work contain several superconducting regions in the graphene thus generating an array of SN junctions. It sould be interesting to study an engineered small device which includes a single N/S interface or an SNS junction. We expect the Josephson effect  to show anomalous behavior as compared to a standard SNS junction.

\vspace{1cm}

\textbf{Data availability}

The data that support the findings of this study are available from
the corresponding author upon request.


\vspace{1cm}

\textbf{Acknowledgments}

We are grateful for technical help from A. Roy and M. Ismail and I. Volotsenko  and for illuminating discussions with
N. Trivedi and J. Ruhman. G.N.D. and A.F were supported by the NSF-BSF US-Israel binational research grant No. 2017677. E.W. and E.S. were supported by the US-Israel
Binational Science Foundation (BSF) grant No. 2016130, NSF-BSF grant No. 2018726, and the Israel Science Foundation (ISF) grant No. 993/19.

\vspace{1cm}

\textbf{Author contributions}

G.N.D. and A.F. designed the experiment and performed the measurements. E.W. and E.S developed the theory and performed the calculations. All authors contributed equally to writing of the manuscript.

\vspace{1cm}
\textbf{Competing interests}

The authors declare no competing interests.

\vspace{1cm}

\section*{Supplementary information - Superconducting Dirac point in proximetized graphene}

\textbf{S1. Sample preperation and measurement technique}\\

\begin{figure}[tbh]
	\begin{center}
		\includegraphics[width=0.48\textwidth]{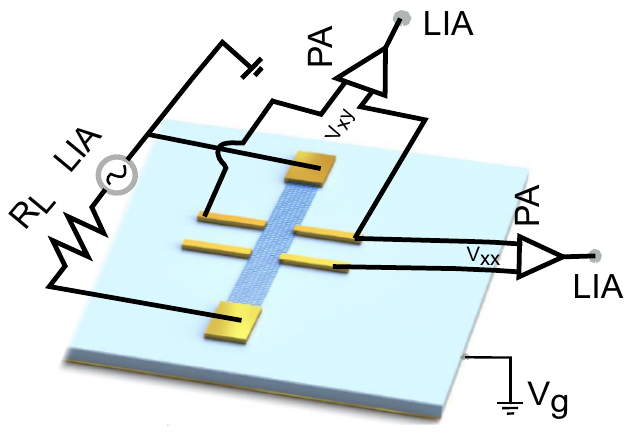}
		\small{\caption{  \textbf{A schematic diagram of the device.} The bare graphene is biased by a constant current and the longitudinal and transverse voltages are amplified by low noise preamplifier (PA- SR552) and fed to lock-in amplifier (SRS 830) in differential mode. The carrier density is modulated by the back gate voltage $V_g$ applied to the contact at the bottom of the Si. \label{device}}}
	\end{center}
\end{figure}

Polycrystalline single layer graphene were prepared by  CVD technique on copper catalyst and then were transferred to SiO$_2$/Si substrate (purchsed from Graphenea company). The samples were patterned into Hall bar geometry by standard e-beam lithography and contacted with Cr/Au leads(5 nm/30 nm). A second stage of lithography was performed for the InO films which were deposited by e-gun evaporation. During the deposition, dry O$_2$ gas was injected into the chamber with partial pressure that allows to tune the system through SIT.

The channel length and width of the sample are 150 $\mu$m and 50 $\mu$m respectively. We evaporate Au on the back side of the Si wafer to act as one metal plate of a capacitor, while graphene acts as the second metallic plate and SiO$_2$ acts as a dielectric material. The carrier density of the graphene device is modulated by changing the gate voltages applied to the back side of Si wafer.
The device structure of SLG along with the electrical connections are shown in Fig. \ref{device}. The electrical transport measurements were performed using standard lockin (SR830) technique. All the electrical measurements were performed in wet He3 system or a 1.5K system with magnetic field of 9T.
\\

\textbf{S2. Hall measurement in SLG}\\

\begin{figure}[tbh]
	\begin{center}
		\includegraphics[width=0.48\textwidth]{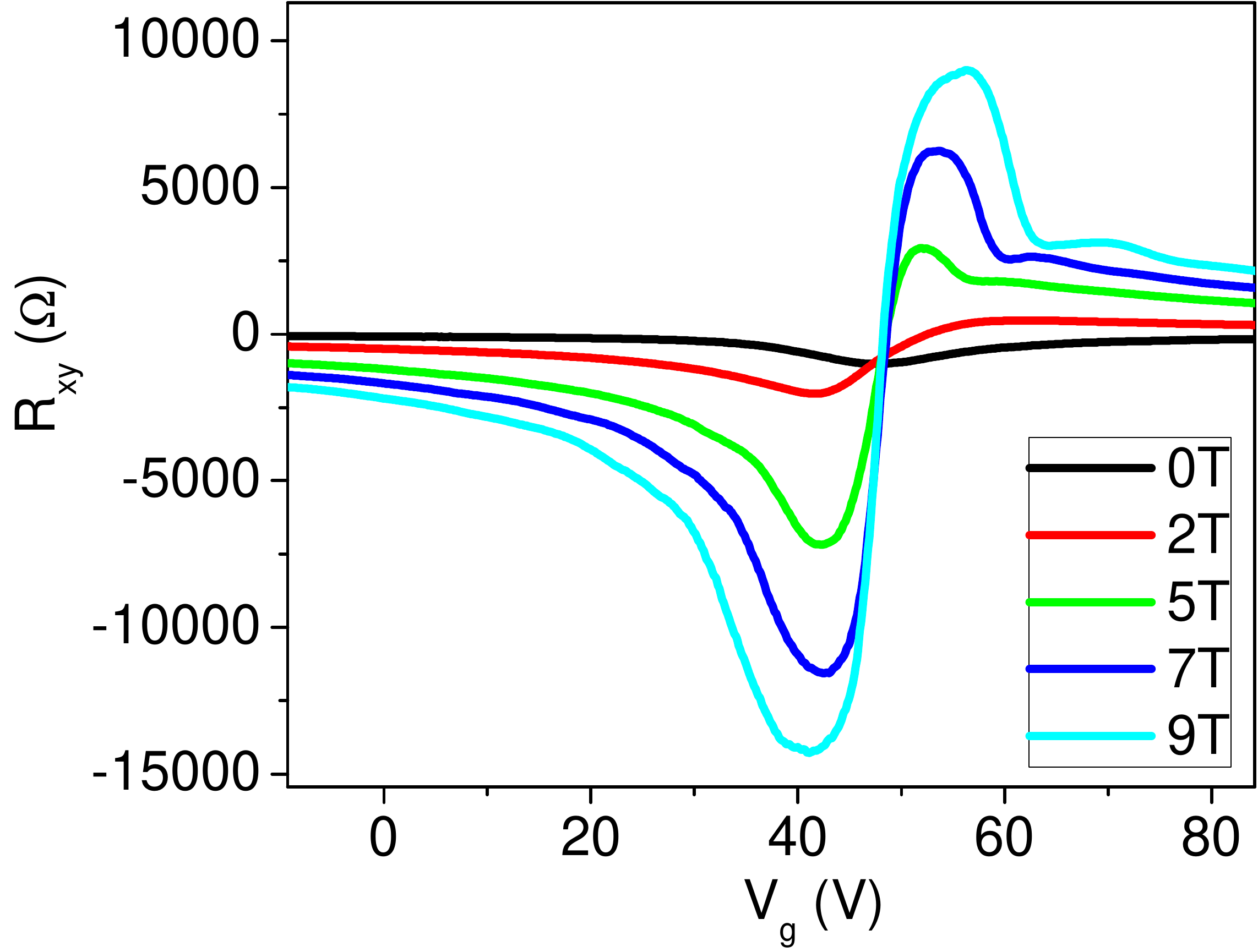}
		\small{\caption{ \textbf{Hall resistance as a function of gate voltage. } Hall resistance $R_{xy}$ as a function of $V_g$ of SLG at $T$= 1.7~K. \label{Rxy_SLG}}}
	\end{center}
\end{figure}

As shown in Fig. 2b in the text, sheet resistance has been scaled with the resistance at the Dirac point (DP). We identified the DP as the charge neutrality point extracted from Hall measurement.  In Fig. \ref{Rxy_SLG}, Hall resistance $R_{xy}$ as a function of $V_g$ for the SLG is shown for different magnetic field and $T = 1.7$ K. The Dirac point occurs at $V_g \sim 50$ V. Similarly, we performed Hall measurement of series of bilayer SLG/a-InO (S1-S4) to identify the DP.\\

\begin{figure*}
	\vspace{0.5cm}
	\includegraphics[width=17cm,height=!]{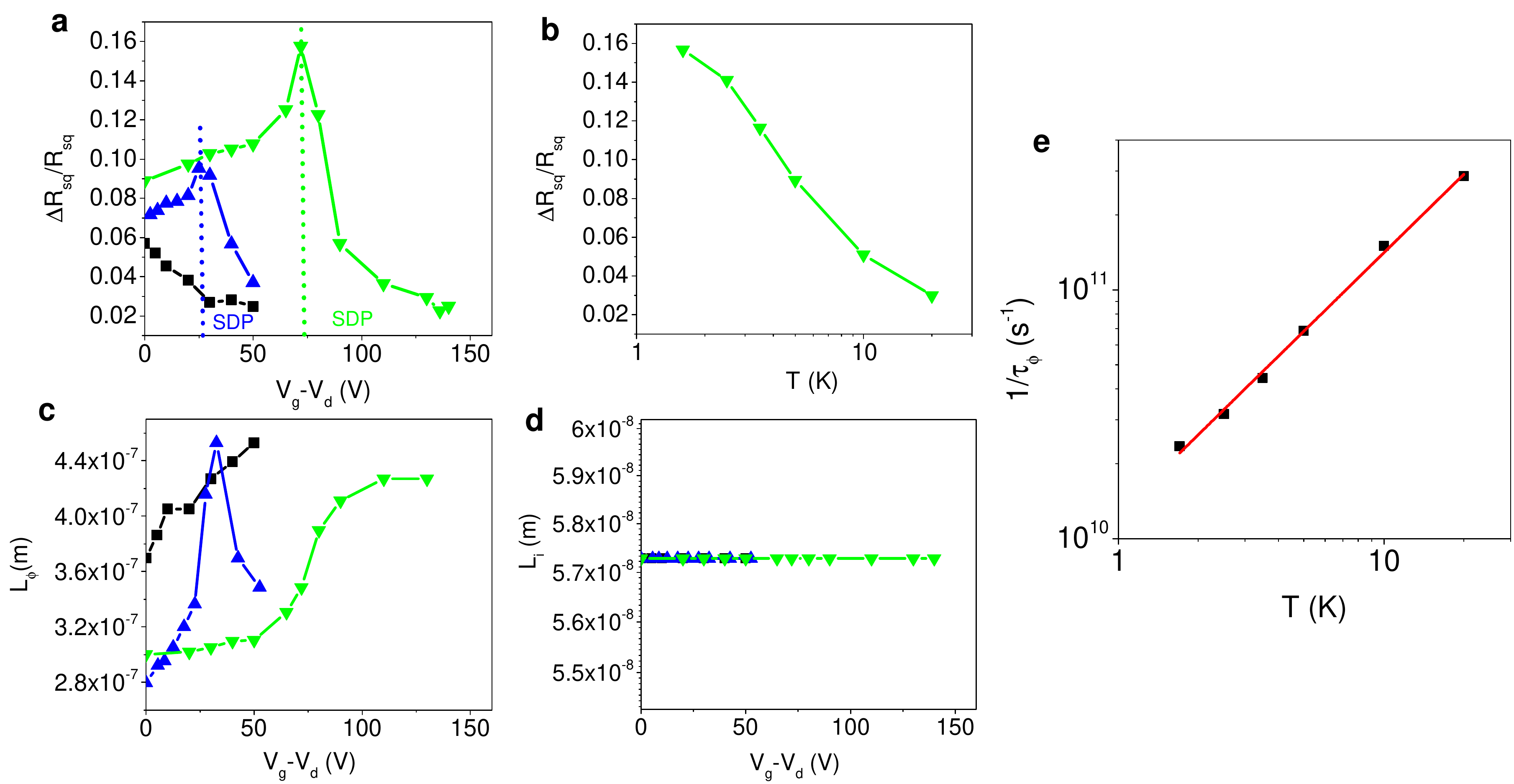}
	\caption{\label{WL1}
		\textbf{Weak localization analysis. a,} Weak localization amplitude as a function of $V_g-V_d$ of SLG (black), S2 (blue) and S4 (green).  \textbf{b,} Weak localization amplitude as a function of $T$ at $V_g-V_d$ = 72 V of S4.  \textbf{c,}   $L_*$ and \textbf{d,} $L_i$  extracted from the fits to Eq. \ref{WAL} as a function of $V_g-V_d$ for the three samples (same color code as \textbf{a,}). \textbf{e,} Plot of $1/\tau_\phi$ as a function of $T$ of S4 (black squares). Linear fit of the data (red line) gives the slope of $\sim 1.04$. Measurents were performed at $T$= 1.7~K. } 
\end{figure*}

\textbf{S2. Weak localization in graphene}\\

WL in graphene is very different than that in conventional metals. Because of the presence of two valleys in K space and the chiral nature of the charge carriers, the interference of carriers is not only sensitive to the inelastic scattering rate, but also, to different elastic scattering processes which may cause decoherence. Charge carriers in graphene acquire a Berry phase of $\pi$ upon completing a closed path thus leading to Weak-antilocalization (WAL) and positive magnetoresistance, MR. However intra-valley chiral-symmetry-breaking scattering as well as time reversal symmetry breaking due to triagonal warping, destroy coherence and thus suppress WAL at an intra-valley scattering rate, $ \tau_*^{-1}$ \cite{tikhonenko2008weak}. Intervalley scattering (at rate $\tau_i^{-1}$) can protect chirality and also nullify the effect of the Berry phase leading to WL and negative MR. Hence, the interplay of the intra-and inter valley scattering processes determine the amplitude and sign of the MR \cite{tikhonenko2009transition}.

At low magnetic field, the MR can be analyzed using the following formula \cite{mccann2006weak}, which depends on several field scales (inelastic $B_\phi$ and elastic $B_i$, $B_*$) in the system 

\begin{align}
	\Delta \rho(B)= &\frac{-e^2 \rho^2}{\pi h}\Big[F\Big(\frac{B}{B_{\phi}}\Big) \notag
	-F\Big(\frac{B}{B_{\phi}+2B_i}\Big) \notag \\
	&-2F\Big(\frac{B}{B_{\phi}+B_*}\Big)\Big].
	\label{WAL}
\end{align}
where $F(z)=\ln(z)+\psi(0.5+\frac{1}{z})$. Here, $\psi$ is the digamma function and $D$ is the Diffusion constant. The phase coherence length $L_\phi$ and elastic intervalley (intravalley) scattering lengths $L_i$($L_*$) can be defined as $L_{\phi,i,*}=\sqrt{\frac{\hbar}{4eB_{\phi,i,*}}}$. 

From Eq. \ref{WAL} one can deduce that the amplitude of  WL is larger the larger are $L_\phi / L_*$ and $L_\phi / L_i$ \ \cite{tikhonenko2009transition}. Indeed, our analysis shows that for all our samples $L_\phi>>L_i\sim L_*$. The data have been fitted with Eq. \ref{WAL} (see Fig. 3a in the main text).

$L_\phi$ for all samples shows a minimum at the DP presumably due to phase breaking scattering events on the boundaries between electron and hole puddles.  Introducing superconducting puddles further reduces $L_\phi$ (Fig. \ref{WL1}c).  On the other hand, at the SDP, $L_\phi$ exhibits a peak or a local maximum. In contrast, $L_*$ shows a sharp minimum at the SDP implying that intervalley scattering is greatly enhanced at this point (see text). Figure \ref{WL1}a shows that the WL magnitude exhibits a sharp peak at the SDP which decreases as increasing temperature (see Fig. \ref{WL1}b). 

In Fig. \ref{WL1}e, we plot the dephasing rate $1/\tau_\phi$ as a function of $T$ of S4 (in log-log scale). The linear fit of the data shows the slope of 1 signifying that electron-electron scattering is the dominant scattering mechanism. \\

\textbf{S4. Theoretical model}\\

When an electron propagates in a conducting material and collides
into a superconductor, there are two fundamentally different reflection
processes -- normal reflection (where the reflected particle is an
electron) and Andreev reflection (AR) whereby a hole is reflected while
an electron pair propagates into the superconductor. Generally, the
electron is reflected as a superposition with coefficients $r,r_{A}$
for reflection and AR correspondingly. The net contribution of the
NS-interface to the conductivity is determined by a balance between
these two processes: normal reflections are reducing it while AR are
enhancing it with respect to the normal-state value. The latter effect
dominates in the case of an ideal NS interface where the normal region
is a standard metal \cite{blonder1982transition}.

In graphene partially overlayed by an s-wave superconductor, the proximity-induced
superconducting pairing field in restricted regions creates a distinct
type of NS interfaces, where the Fermi energy in both the N and S
sides is tunable across the Dirac points by gating. Notably, since
pairing is introduced between electrons in opposite valleys, the AR
are inter-valley processes. In contrast, normal reflections from an
NS interface contribute to intra-valley processes, which in pristine
graphene are relatively suppressed. As we show below, the balance
between these two types of processes changes dramatically as a function
of gate voltage.

\subsection*{S4.1. Derivation of $r$ and $r_{A}$}

To model the system, we consider a clean NS interface parellel to
the $y$-direction where the half-plane $x>0$ is normal (N) and $x<0$
is superconducting (S). Such an interface is described by three parameters
-- the Fermi energy $E_{F}$ (defined in the normal side), the superconducting
gap in the S side $\Delta_{0}$, and the electrical potential difference
$U_{0}$ which effectively shifts the Fermi energy in the S
side to $E_{F}^{\prime}=E_{F}+U_{0}$. While $E_{F}$ (determined
by the gate voltage) is uniform, we adopt a step-function model for
the other two \cite{beenakker2006specular}:
\begin{widetext}
	\begin{equation}
		\label{step_Delta_U}
		\Delta\left(\vec{r}\right)=\begin{cases}
			\Delta_{0}e^{i\phi} & x<0\\
			0 & x>0
		\end{cases}
		\; ,\quad
		U\left(\vec{r}\right)=\begin{cases}
			-U_{0} & x<0\\
			0 & x>0
		\end{cases}\; .
	\end{equation}
\end{widetext}
These piecewise-constant potentials are then incorporated in the Dirac-Bogoliubov-de
Gennes equation (DBdG) \cite{beenakker2006specular} describing the excitations in Nambu space:
\begin{widetext}
	\begin{equation}
		\begin{pmatrix}H_{\pm}+U(x)-E_{F} & \Delta(x)\\
			\Delta^{*}(x) & E_{F}-U(x)-H_{\pm}
		\end{pmatrix}\begin{pmatrix}u\\
			v
		\end{pmatrix}=\varepsilon\begin{pmatrix}u\\
			v
		\end{pmatrix}
		\label{DBdG}
	\end{equation}
\end{widetext}
where $u$ is the electron pseudo-spinor for one valley ($\pm\rightleftarrows K,K^{\prime}$)
and $v$ is the hole pseudo-spinor in the other valley; $H_{\pm}=v_F(p_x\sigma_x\pm p_y\sigma_y)$ where $\sigma_i$ are the Pauli matrices.
In view of Eq. (\ref{step_Delta_U}) for $U$, one identifies two distinct Dirac points accessible by the tuning of $E_F$: the regular
DP $E_{F}=0$, where the N side is in its charge neutrality
point (CNP), and the superconducting Dirac
point (SDP) corresponding to $E_{F}^{\prime}=E_{F}+U_{0}=0$ where the S
side is in its CNP.

To calculate the rates of reflection and AR at the NS interface, we first solve Eq. (\ref{DBdG}) to get the quasi-particle states in each of the two sides, and then require continuity along the interface $x=0$. The excitation energy $\varepsilon$ and the momentum in the $y$-direction (with corresponding wave-vector $q$) are both good quantum
numbers, restricting the basis states considered below.
In the normal side we obtain four states:
\begin{align}
	\psi^{e+} & =\frac{e^{iqy+ikx}}{\sqrt{\cos\alpha}}\begin{pmatrix}e^{-\nicefrac{i\alpha}{2}}\\
		e^{\nicefrac{i\alpha}{2}}\\
		0\\
		0
	\end{pmatrix}\\
	\psi^{e-} & =\frac{e^{iqy-ikx}}{\sqrt{\cos\alpha}}\begin{pmatrix}e^{\nicefrac{i\alpha}{2}}\\
		e^{-\nicefrac{i\alpha}{2}}\\
		0\\
		0
	\end{pmatrix}\\
	\psi^{h+} & =\frac{e^{iqy+ik^{\prime}x}}{\sqrt{\cos\alpha^{\prime}}}\begin{pmatrix}0\\
		0\\
		e^{-\nicefrac{i\alpha^{\prime}}{2}}\\
		e^{\nicefrac{i\alpha^{\prime}}{2}}
	\end{pmatrix}\\
	\psi^{h-} & =\frac{e^{iqy-ik^{\prime}x}}{\sqrt{\cos\alpha^{\prime}}}\begin{pmatrix}0\\
		0\\
		e^{\nicefrac{i\alpha^{\prime}}{2}}\\
		e^{-\nicefrac{i\alpha^{\prime}}{2}}
	\end{pmatrix}
\end{align}
with the definitions
\begin{align}
	\alpha & =\arcsin\left(\frac{\hbar v_Fq}{\varepsilon+E_{F}}\right)\\
	\alpha^{\prime} & =\arcsin\left(\frac{\hbar v_Fq}{\varepsilon-E_{F}}\right)\\
	k & =\frac{\varepsilon+E_{F}}{\hbar v_F}\cos\alpha\\
	k^{\prime} & =\frac{\varepsilon-E_{F}}{\hbar v_F}\cos\alpha^{\prime}\; .
\end{align}
Here $\alpha$ is the angle of incidence (with respect to the $x$-axis) for an incoming electron, and $k$ the wave-vector component
perpendicular to the interface; the angle of Andreev reflection
is $\alpha^{\prime}$ and the perpendicular wave-vector component
is $k^{\prime}$. $\psi^{e/h+}$ is an electron/hole propagating away from
the interface, and $\psi^{e/h-}$ propagate towards it. The cosine
factors ensure the states carry a unit particle current.

In the S side, the DBdG equation Eq. (\ref{DBdG}) with $\Delta\neq 0$ yields superpositions of holes and electrons. In
the most general case, there are four solutions; for low excitation energy $\varepsilon<\Delta_0$, two of them
are decaying waves and thus satisfy the boundary condition at infinity. The form of the solutions depends also on the sign of $E_{F}^{\prime}\equiv E_{F}+U_{0}$.
For positive $E_{F}^{\prime}$ (electronically doped states), we get
\begin{align}
	\psi^{S-} & =e^{iqy+ik_{+}x}\begin{pmatrix}e^{-i\beta}\\
		e^{-i\beta+i\gamma_{+}}\\
		e^{-i\phi}\\
		e^{i\gamma_{+}-i\phi}
	\end{pmatrix},\\
	\psi^{S+} & =e^{iqy+ik_{-}x}\begin{pmatrix}e^{i\beta}\\
		e^{i\beta+i\gamma_{-}}\\
		e^{-i\phi}\\
		e^{i\gamma_{-}-i\phi}
	\end{pmatrix},
\end{align}
and for negative $E_{F}^{\prime}$ (hole doped states)
\begin{align}
	\psi^{S-} & =e^{iqy+ik_{+}x}\begin{pmatrix}e^{-i\beta}\\
		-e^{-i\gamma_{+}-i\beta}\\
		e^{-i\phi}\\
		-e^{-i\gamma_{+}-i\phi}
	\end{pmatrix},\\
	\psi^{S-} & =e^{iqy+ik_{+}x}\begin{pmatrix}e^{i\beta}\\
		-e^{-i\gamma_{-}+i\beta}\\
		e^{-i\phi}\\
		-e^{-i\gamma_{-}-i\phi}
	\end{pmatrix}\; .
\end{align}
The parameters -- the (generally complex) wave vectors $k_{\pm}$,
$\beta$ and $\gamma_{\pm}$ -- are defined by
\begin{widetext}
	\begin{align}
		\beta & =\arccos\left(\nicefrac{\varepsilon}{\Delta_{0}}\right)\\
		\gamma_{\pm} & =\frac{1}{i}\log\left(\frac{\pm\sqrt{E_{F}^{\prime2}+\varepsilon^{2}-\Delta_{0}^{2}-(\hbar v_Fq)^{2}\pm2i\left|E_{F}^{\prime}\right|\sqrt{\Delta_{0}^{2}-\varepsilon^{2}}}+i\hbar v_Fq}{\sqrt{E_{F}^{\prime2}+
				\varepsilon^{2}-\Delta_{0}^{2}\pm2i\left|E_{F}^{\prime}\right|\sqrt{\Delta_{0}^{2}-\varepsilon^{2}}}}\right)\label{eq:gammaExpression}\\
		k_{\pm} & =\pm\frac{\sqrt{E_{F}^{\prime2}-(\hbar v_Fq)^{2}+\varepsilon^{2}-\Delta_{0}^{2}\pm2\left|E_{F}^{\prime}\right|\sqrt{\varepsilon^{2}-\Delta_{0}^{2}}}}{\hbar v_F}\label{eq:kExpression}
	\end{align}
\end{widetext}

\begin{figure*}
	\begin{centering}
		\includegraphics[scale=0.5]{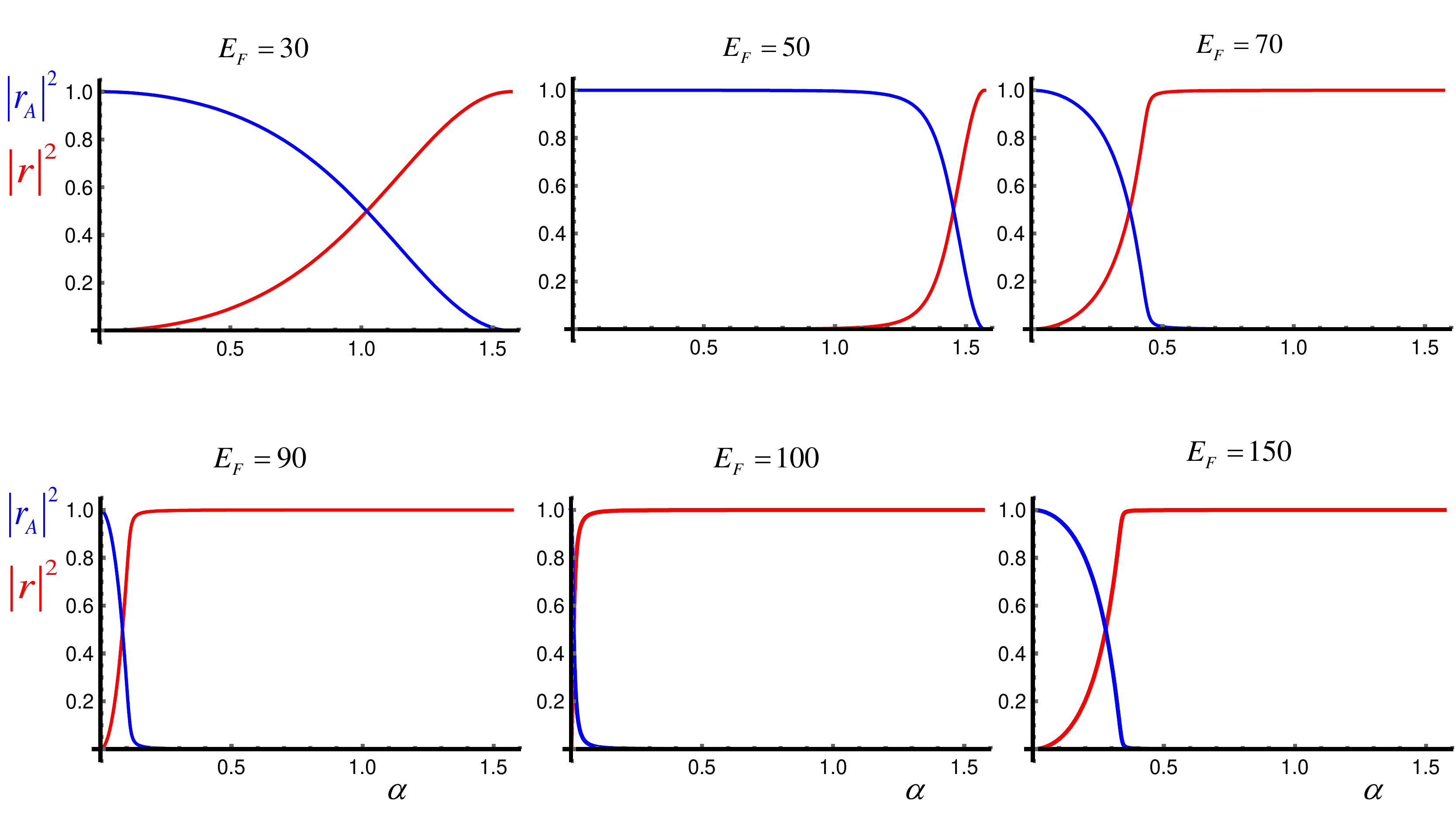}
		\par\end{centering}
	\caption{\label{fig:ReflectionRates}\textbf{Normal and Andreev reflection rates:} The reflection rates $|r|^2$ and $|r_A|^2$ (red and blue curves respectively) as a function of incidence angle for different value of $E_F$ (given in units of $\Delta_0$). Here $U_0=-100$, so that the top middle panel corresponds to the perfect matching point $E_F=\left|E_{F}^{\prime}\right|$, and the bottom middle panel to the SDP $E_{F}^{\prime}=0$ (see text).}
\end{figure*}

We now consider a particular linear combination describing an in-going electron in N and no in-going holes,
and impose continuity at the NS-interface. This yields the equation
\[
\psi^{e-}+r\psi^{e+}+r_{A}\psi^{h+}=a\psi^{S+}+b\psi^{S-}
\]
at $x=0$, where $r,r_{A}$ are the reflection and Andreev
reflection rates and $a,b$ the amplitudes of the modes in the superconductor.
The solution for $r,r_{A}$ as functions of $\alpha$ and $E_F$ is presented graphically in Fig. \ref{fig:ReflectionRates}.
Note that due to the decaying nature of states in the S region, the unitarity condition $\left|r\right|^{2}+\left|r_{A}\right|^{2}=1$ is obeyed.
For arbitrary $E_F$, perfect AR ($r_{A}=1$) occur at normal incidence to the interface ($\alpha=0$), and is being traded by perfect normal reflection at higher incidence angles. The most prominent effect of tuning $E_F$ is the shift of the threshold angle where this switching occurs.
The general exact formula describing this behavior is cumbersome; however, a relatively simple
analytical result can be obtained in some limits, providing insight on the evolution of these reflection coefficients with $E_F$ exhibited in the figure. Focusing on the low-energy limit $\varepsilon\rightarrow 0$ (dictating the d.c. conductance),
we consider below two different limits: $\Delta_{0}\ll E_{F}^{\prime}$
and $E_{F}^{\prime}\ll\Delta_{0}$.

Away from either of the Dirac points, $\Delta_{0}\ll E_{F}^{\prime},E_{F}$.
In this limit, although superconductivity is accounted for by allowing
electron-hole conversion, $\Delta_{0}$ can be neglected in Eqs. (\ref{eq:gammaExpression}),(\ref{eq:kExpression}).
The behavior of reflection rates is characterized by a critical angle
corresponding to $q=\frac{E_{F}^{\prime}}{\hbar v_F}$. The reflection
rates in this limit are
\begin{equation}
	\label{r_EFlarge}
	\left|r\right|=\begin{cases}
		\frac{\left|\left|E_{F}^{\prime}\right|-E_{F}\right|\sin\alpha}{\left|\left|E_{F}^{\prime}\right|-E_{F}\sin^{2}\alpha\right|} & \sin\alpha<\frac{\left|E_{F}^{\prime}\right|}{E_{F}}\\
		1 & \sin\alpha>\frac{\left|E_{F}^{\prime}\right|}{E_{F}}
	\end{cases}\; ,
\end{equation}
\begin{equation}
	\label{rA_EFlarge}
	\left|r_{A}\right|=\begin{cases}
		\frac{\sqrt{E_{F}^{\prime2}-E_{F}^{2}\sin^{2}\alpha}}{\left|\left|E_{F}^{\prime}\right|-E_{F}\sin^{2}\alpha\right|} & \sin\alpha<\frac{\left|E_{F}^{\prime}\right|}{E_{F}}\\
		0 & \sin\alpha>\frac{\left|E_{F}^{\prime}\right|}{E_{F}}
	\end{cases}\; .
\end{equation}
These expressions are in good agreement with the exact result presented in Fig. \ref{fig:ReflectionRates}, except for the bottom middle panel (see discussion below). They reflect the general behavior seen in the experiment: as $E_F$ is tuned upwards, one first observes
an increase in conductance until $E_{F}=-\frac{U_{0}}{2}$. In this point, corresponding to $|E_{F}^{\prime}|=E_F$, there is perfect AR and no regular reflection at almost any $\alpha$ (i.e. $\alpha_c\approx \pi/2$). A further increase of $E_F$ away from this ideal point
reveals a very sharp decrease in conductance towards the region of the
SDP $E_{F}=-U_{0}$, where regular reflections dominate over AR at almost any $\alpha$. Beyond this conductance minimum point, the relative weight of AR is gradually recovered.

In the close vicinity of the SDP ($|E_{F}^{\prime}|\ll\Delta_{0}$), the above approximation fails and the finite value of $\Delta_{0}$ becomes crucial. We then use the opposite approximation, $E_{F}^{\prime}\approx 0$. The resulting reflection coefficients are given by
\begin{align}
	\label{r_SDP}
	r & =\sin\alpha\left[\frac{E_{F}\cos\alpha}{\sqrt{\Delta_{0}^{2}+E_{F}^{2}\sin^{2}\alpha}}-i\right]\\
	r_{A} & =-ie^{-i\phi}\cos\alpha\frac{\Delta_{0}}{\sqrt{\Delta_{0}^{2}+E_{F}^{2}\sin^{2}\alpha}}\; .
\end{align}
Notably, the corresponding reflection rates
\begin{align}
	\left|r\left(\alpha\right)\right| & =\frac{\sin\alpha\sqrt{\Delta_{0}^{2}+E_{F}^{2}}}{\sqrt{\Delta^{2}+E_{F}^{2}\sin^{2}\alpha}}\; ,\\
	\left|r_{A}\left(\alpha\right)\right| & =\frac{\cos\alpha\Delta_{0}}{\sqrt{\Delta_{0}^{2}+E_{F}^{2}\sin^{2}\alpha}}
	\label{rA_SDP}
\end{align}
exhibit a threshold dependence on $\alpha$ at a finite critical incidence angle $\alpha_c\approx \arcsin(\Delta_0/E_F)$. This implies that AR still prevails for normal incident electrons, within a range of angles $\alpha\lesssim \alpha_c$; they correspond to incoming states with parallel wave-vector component $q\lesssim 1/\xi$, where $\xi\sim\frac{\Delta_{0}}{\hbar v_{F}}$ is the superconducting coherence length.

\subsection*{S4.2. Conductance}

We next evaluate the conductance across
the NS-intercade, using the Blonder-Tinkham-Klapwijk formula \cite{blonder1982transition} which averages over contributions from all the incidence angles $\alpha$:
\begin{widetext}
	\begin{equation}
		\label{dIdVFormula}
		\sigma_s\equiv\frac{dI}{dV}=\sigma_{N}\left(V\right)\int_{0}^{\nicefrac{\pi}{2}}d\alpha\left(1-\left|r\left(eV,\alpha\right)\right|^{2}+\left|r_{A}
		\left(eV,\alpha\right)\right|^{2}\right)\cos\alpha
	\end{equation}
\end{widetext}
where $\sigma_{N}$ is the normal conductance. Having evaluated $r,r_A$ from the general solution of Eq. (\ref{DBdG}) at arbitrary $\varepsilon =eV$, this yields the differential conductance as a function of $V$ and gate voltage parametrized by $E_F$, for a give choice of $U_0$ and $\Delta_0$. The result for the d.c. limit $V=0$ is presented in Fig. 4h of the main text. The functional dependence on $E_F$ (which exhibits distinct features over energy scales $\sim U_0\gg\Delta_0$) is captured to a good approximation by an analytical expressions, obtained by substituting Eqs. (\ref{r_EFlarge}), (\ref{rA_EFlarge}) in (\ref{dIdVFormula}):
\begin{widetext}
	\begin{equation}
		\frac{\sigma_s}{\sigma_{N}}  =
		\begin{cases}
			-1+\frac{3\left|E_{F}^{\prime}\right|}{E_{F}}+\frac{1}{2}\left(1-\frac{\left|E_{F}^{\prime}\right|}{E_{F}}\right)^{2}\sqrt{\frac{E_{F}}{\left|E_{F}^{\prime}\right|}}\ln\left(\frac{1+\sqrt{\nicefrac{\left|E_{F}^{\prime}\right|}{E_{F}}}}{1-\sqrt{\nicefrac{\left|E_{F}^{\prime}\right|}{E_{F}}}}\right) & E_{F}>\left|E_{F}^{\prime}\right|\left(E_{F}>\frac{\left|U_{0}\right|}{2}\right)\\
			3-\frac{\left|E_{F}^{\prime}\right|}{E_{F}}+\frac{1}{2}\left(\frac{\left|E_{F}^{\prime}\right|}{E_{F}}-1\right)^{2}
			\sqrt{\frac{E_{F}}{\left|E_{F}^{\prime}\right|}}\ln\left(\frac{1+\sqrt{\nicefrac{\left|E_{F}^{\prime}\right|}{E_{F}}}}{1-\sqrt{\nicefrac{\left|E_{F}^{\prime}\right|}{E_{F}}}}\right) & E_{F}<\left|E_{F}^{\prime}\right|\left(E_{F}<\frac{\left|U_{0}\right|}{2}\right)
		\end{cases} \; .
	\end{equation}
\end{widetext}
This result is valid in the limit
$\Delta_{0}\ll E_{F},\left|E_{F}^{\prime}\right|$.
In the vicinity of the SDP, we use Eqs. (\ref{r_SDP}), (\ref{rA_SDP}) obtained the opposite approximation
$\left|E_{F}^{\prime}\right|\ll\Delta_{0}$, to get the minimum of conductance at $E_{F}^{\prime}=0$:
\begin{equation}
	\frac{\sigma_s}{\sigma_{N}}=\frac{\pi\Delta_{0}}{E_{F}}\; .
\end{equation}
Once again, this result manifests the contribution from AR of electrons impinging at the interface within the narrow range of deviations from the forward direction limited by superconducting coherence. 


\vspace{1cm}

\textbf{S5. dI/dV analysis}\\

\begin{figure}[tbh]
	\begin{center}
		\includegraphics[width=0.48\textwidth]{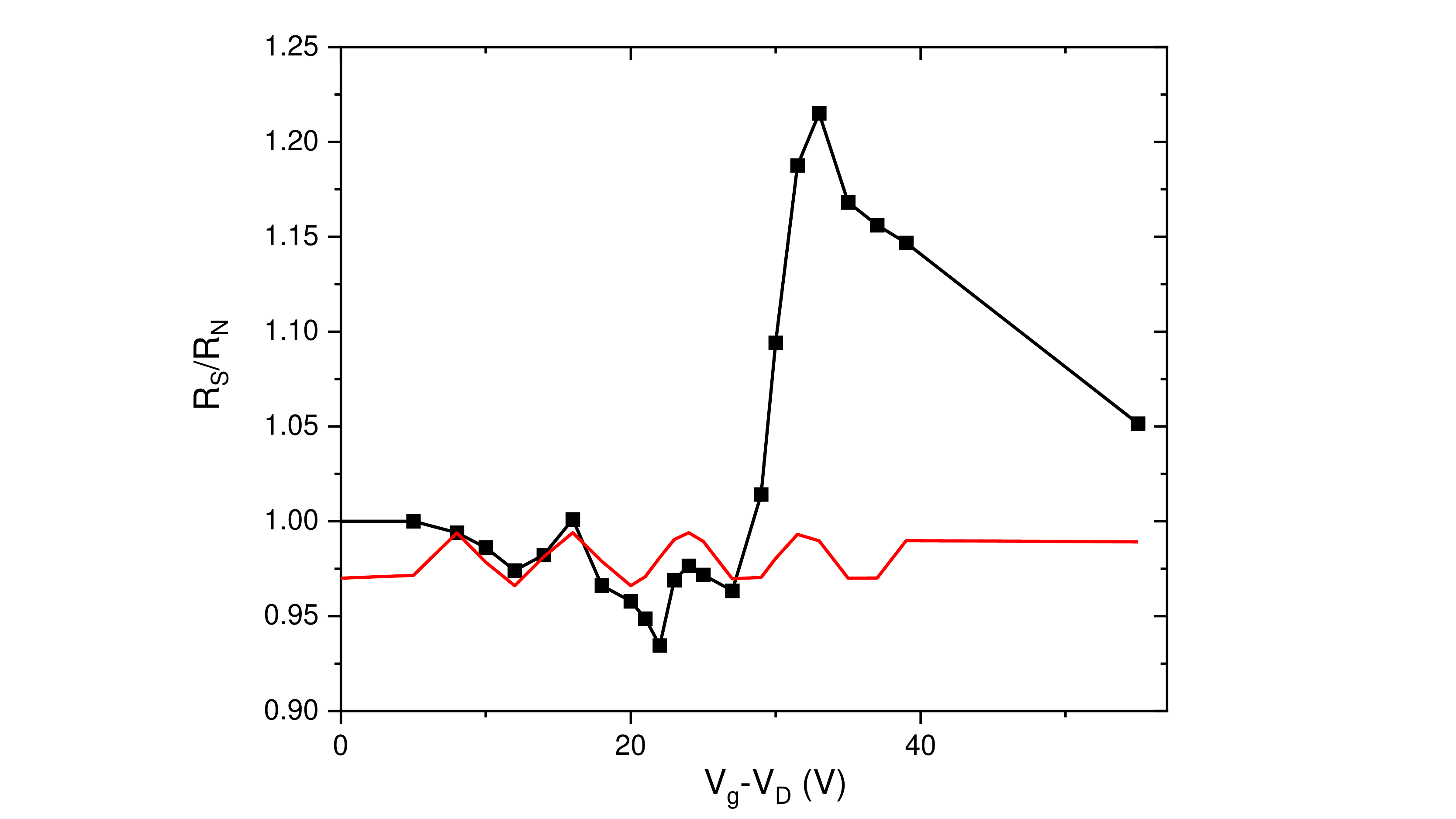}
		\small{\caption{ \textbf{dI-dV analysis}.  Differential resistance of S3 at 6T normalized to that of SLG (black curve) and the phenomenological formula of Eq. \ref{sdh} (red curve). Fig. 4g of the main text is obtained from their subtraction. \label{6T}}}
	\end{center}
\end{figure}

In order to avoid WL effects the dI/dV measurements were performed at $B=6$T. For low values of $V_g$ one can not avoid traces of quantum oscillations due to the Shubnikov de Haas effect. For obtaining Fig. 4g we assume a phenomenological function for these osculations:

\begin{equation}
	\label{sdh}  
	R=A+B*sin(2 \pi * \nu/4)
\end{equation}

where A and B are fitting parameters and $\nu$ is the calculated filling factor. 
and subtract it from the raw data at B=6T. The raw data and function are plotted in Fig. \ref{6T}.


\end{document}